\documentclass[review,number,sort&compress]{elsarticle}
\usepackage{amssymb}
\usepackage{amsmath}
\usepackage{multirow}
\usepackage{booktabs}
\usepackage{lscape}
\usepackage{graphicx}
\usepackage{threeparttable}
\usepackage{footnote}
\usepackage[a4paper]{geometry}
\usepackage[]{natbib}
\usepackage{hyperref}
\usepackage{rotating}
\begin{document}

\begin{frontmatter}

\title{In-gas-cell laser ionization spectroscopy in the vicinity of $^{100}$Sn:
Magnetic moments and mean-square charge radii of N$=$50-54 Ag}

\author[KUL]{R. Ferrer\corref{cor1}}
%\ead{Rafael.Ferrer@fys.kuleuven.be, Tel.:+32 16327271, Fax:+32 16327985}
\author[KUL]{N.~Bree}
\author[KUL]{T.E.~Cocolios\fnref{fn1}}
\author[KUL]{I.G.~Darby}
\author[KUL]{H.~De~Witte}
\author[KUL]{W.~Dexters}
\author[KUL]{J.~Diriken}
\author[KUL]{J.~Elseviers}
\author[IPN]{S.~Franchoo}
\author[KUL]{M.~Huyse}
\author[KUL,SCK]{N.~Kesteloot}
\author[KUL]{Yu.~Kudryavtsev}
\author[KUL]{D.~Pauwels}
\author[KUL]{D.~Radulov}
\author[KUL]{T.~Roger\fnref{fn2}}
\author[GANIL]{H.~Savajols}
\author[KUL]{P.~Van~Duppen}
\author[KUL]{and \,M.~Venhart\fnref{fn3}}

\address[KUL]{KU Leuven, Instituut voor Kern- en Stralingsfysica, Celestijnenlaan 200D, B-3001 Leuven, Belgium}
\address[IPN]{Institut de Physique Nucl\'{e}aire (IPN) d'Orsay, 91406 Orsay, Cedex, France}
\address[GANIL]{GANIL, CEA/DSM-CNRS/IN2P3, B.P. 55027, 14076 Caen, France}
\address[SCK]{SCK-CEN, Studiecentrum voor Kernenergie - Centre d'Etude de l'énergie Nucléaire, Mol B-2400, Belgium}
\cortext[cor1]{Corresponding author}
\fntext[fn1]{Present address: School of Physics and Astronomy, University of Manchester, Brunswick Street, M13 9PL Manchester, UK}
\fntext[fn2]{Present address: GANIL, CEA/DSM-CNRS/IN2P3, B.P. 55027, 14076 Caen, France}
\fntext[fn3]{Present address:  Institute of Physics, Slovak Academy of Science, Slovakia}

\begin{abstract}
In-gas-cell laser ionization spectroscopy studies on the neutron deficient $^{97-101}$Ag isotopes have been performed with the LISOL setup.
Magnetic dipole moments and mean-square charge radii have been determined for the first time with the exception of $^{101}$Ag, which was found in
good agreement with previous experimental values. The reported results allow tentatively assigning the spin of $^{97,99}$Ag to $\frac{9}{2}$ and confirming the presence of an isomeric state in these two isotopes, whose collapsed hyperfine structure suggests a spin of $\frac{1}{2}$. The effect of the N$=$50 shell closure is not only manifested in the magnetic moments but also in the evolution of the mean-square charge radii of the isotopes investigated, in accordance with the spherical droplet model predictions.

\end{abstract}

\begin{keyword}
% Resonance ionization \sep CHECK  \sep laser ion source  \sep gas cell \sep

\PACS 29.25.Ni \sep 29.25.Rm \sep 41.85.Ar

\end{keyword}

\end{frontmatter}

%\linenumbers

\section{Introduction}

The shell structure of the doubly-magic (N$=$Z$=$50) nucleus $^{100}$Sn, and that of the nuclei in its vicinity, represent an ideal workbench to gain a better understanding in nuclear shell structure far from stability as well as to address important astrophysical processes related, e.g., to the nucleosynthesis rp-process or to the knowledge of the Gamow-Teller transitions, which govern electron-capture during the final stage of massive stars.

Recent experiments on the decay of $^{100}$Sn have determined the smallest known log(ft) value--greatest Gamow-Teller strength--of any nuclear $\beta$-decay yet observed, providing evident indications for the robustness of N$=$Z$=$50 shell closures, and in addition, allowing a stringent test of large scale shell-model calculations in this region \cite{Hin12}.

Near $^{100}$Sn, three-quasi-particle cluster states  have provided explanation for the location of low-lying states in the neutron deficient silver isotopes and offered a possible explanation for the low d$_{5/2}$-g$_{7/2}$ neutron separation energy in $^{103}$Sn \cite{Res02}. An alternative solution  to explain the location of these states has recently been suggested being instead due to enhanced pairing in the g$_{7/2}$ orbital \cite{Dar10}.

In the neutron-deficient silver isotopes, nuclear-decay spectroscopy studies have been performed down to the isotope $^{94}$Ag, at the proton drip line \cite{Muk06}. However, ground- and isomeric-state properties such as spins, moments and charge radii have been determined in laser spectroscopy experiments only for the heavier silver isotopes down to $^{101}$Ag \cite{Din89}, still four neutrons away from the magic N$=$50 shell closure, while Penning-trap mass spectrometry has provided precise mass measurements down to $^{98}$Ag \cite{Her11}. The magnetic moments of the even-A $^{102-110}$Ag isotopes are discussed in view of the competition between the ($\pi$g$_{9/2}$)$^{-3}_{7/2^{+}}$ and ($\pi$g$_{9/2}$)$^{-3}_{9/2^{+}}$ proton groups coupling to ($\nu$d$_{5/2}\nu$g$_{7/2}$)$_{5/2^{+}}$ neutron configurations \cite{Gol10}.

%Interesting results on the nuclear-coupling scheme within the silver isotopes $^{102-110}$Ag are discussed in \cite{Gol10}.

A significant improvement of the experimental possibilities using radioactive ion beams (RIB) was made possible by the implementation of in-source resonance laser ionization, which is characterized by the beam purity enhancement due to its selective ionization nature \cite{Alk89, Mis93, Ver94}. In recent years, this technique has further evolved and in-source resonance laser ionization spectroscopy has been accomplished as reported e.g. in \cite{Coc11}. Recent modifications to a gas-cell-based laser ion source \cite{Yur09} have enabled resonance laser ionization spectroscopy to be performed with spectral profiles that allow nuclear structure information to be extracted. First proof of principle of this method using light-ion reactions has recently been demonstrated \cite{Coc09,Coc10}.

In this paper we extended the applicability of this technique, using heavy-ion induced reactions, to study the lighter silver isotopes from $^{101}$Ag down to $^{97}$Ag, at the N$=$50 shell closure. Magnetic dipole moments and mean-square charge radii of these isotopes have been determined for the first time, with the exception of $^{101}$Ag. A gradual evolution of the underlying nuclear structure towards the N$=$50 shell closure is observed in the results reported.

\section{Experimental Procedure}
The experiments were performed at the Leuven Isotope Separator On-Line (LISOL) facility \cite{Yur03} of the \emph{Centre de Recherche du Cyclotron} in Louvain-La-Neuve, Belgium \cite{Huy11}. The silver nuclei were produced in heavy-ion fusion evaporation reactions of $^{92}$Mo($^{14}$N,2pxn)$^{104-x}$Ag (for $x=$ 3, 4, 5) and $^{nat}$Zn($^{36}$Ar,pxn)$^{101-x}$Ag (for $x=$ 3, 4) with a center of target beam energy of 130~MeV and 125~MeV, respectively. The primary nitrogen or argon beams entered the dual-chamber gas cell impinging on a molybdenum ($>$97\% enriched, 3.29~mg/cm$^{2}$ at an inclination of 55$^{\circ}$ relative to the incoming beam) or zinc ($^{64}$Zn $>$98\% enriched, 3.57~mg/cm$^{2}$ 70$^{\circ}$inclination) foil target.

\begin{figure}
\begin{center}
\includegraphics[scale=0.45]{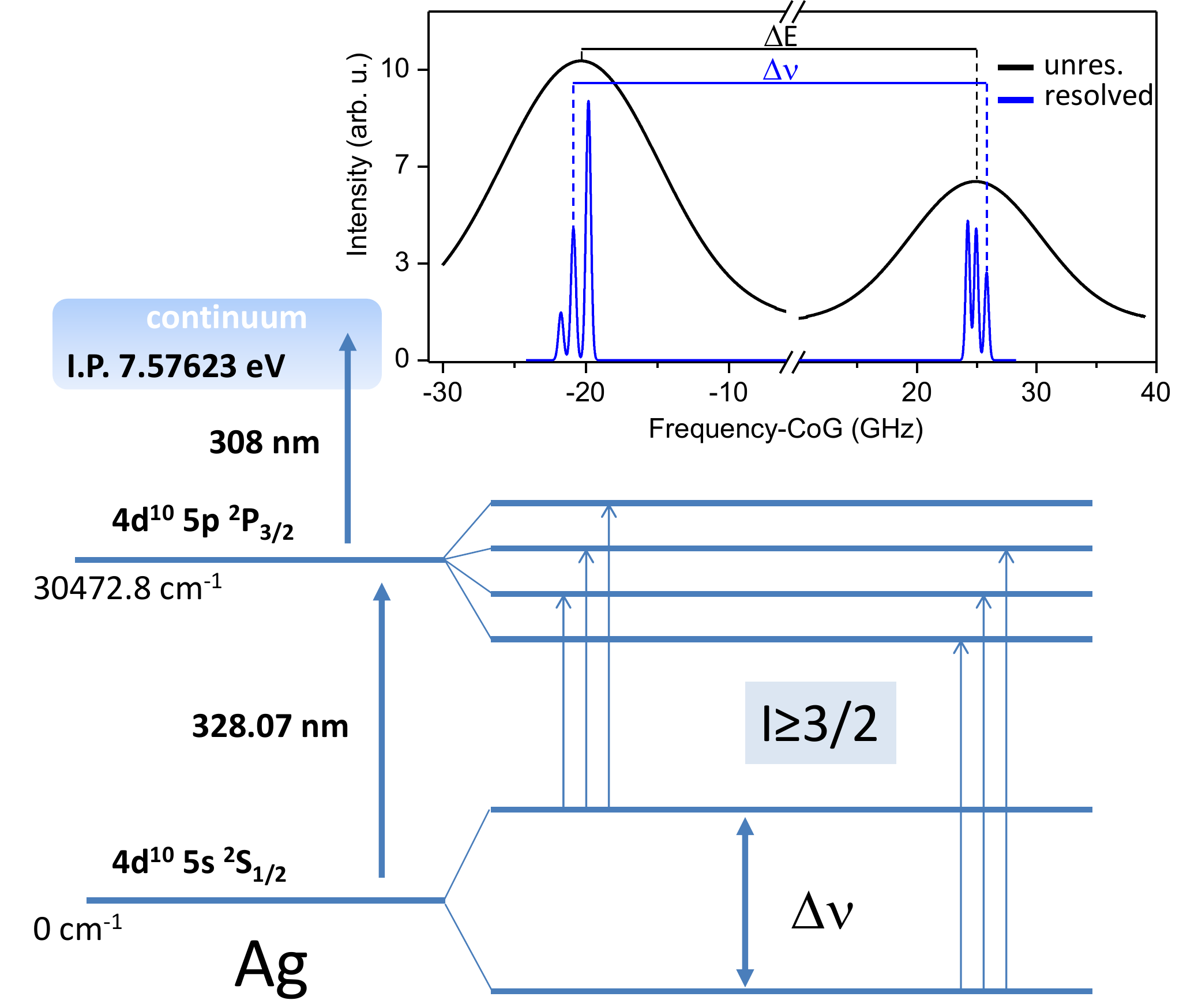} \caption{Ionization scheme employed to ionize silver in these experiments and the expected hyperfine splitting when coupled to nuclear spins I$\geq$3$/$2. The graph illustrates laser frequency scans near the Center of Gravity (CoG) of the transition with unresolved and resolved hyperfine structures.}\label{ion:sch}
\end{center}
\end{figure}

Recoiling fusion-evaporation reaction products were thermalized and neutralized within the gas cell filled with high-purity argon at an operating pressure of 390(5) or 520(5) mbar for the nitrogen or argon beams, respectively. The reaction products were resonantly laser ionized \cite{Yur96}, extracted via a 0.5 mm diameter exit hole and transported by a radio-frequency sextupole ion guide (SPIG) up to the mass separator, where the beam was isotopically separated according to its mass-to-charge ratio A/Q. The resulting isobaric RIB was subsequently deposited on the aluminized mylar tape of a tape station, comprising of three plastic scintillator detectors for $\beta$-decay counting and two single-crystal co-axial HPGe detectors for $\gamma$-ray detection, all arranged in close geometry around the deposition point at the focal plane (this setup is similar to that described in \cite{Paw09}).

\subsection{Laser Ionization}
The LISOL laser system has been extensively described in a previous publication \cite{Yur96}. In these experiments laser ionization was carried out following the two-step two-color ionization scheme shown in Fig. \ref{ion:sch}. A tunable dye laser pumped by a XeCl excimer laser was utilized for the first step excitation. The ionization step, using a non-resonant transition to the continuum, was performed by the direct output of a second XeCl excimer laser at 308~nm. The lasers were operated synchronously with a pulse length of 15~ns and a repetition rate of 200~Hz.

Laser spectroscopy was performed by scanning the transition from the 4d$^{10}$ 5s$^2$ S$_{1/2}$  ground state to the 4d$^{10}$ 5p$^2$ P$_{3/2}$  excited state  at 328.16~nm. To minimize the laser bandwidth during the frequency scans an etalon (5 mm air-gap spaced) was inserted in the resonator cavity of the dye laser as an additional dispersive element. During off-line measurements, the laser radiation  was analyzed using a Fabry-P\'{e}rot interferometer revealing three to four longitudinal oscillation modes in the dye laser resonator. The observed modes (FWHM$=$150 MHz), separated from one another by 400 MHz, resulted in a linewidth for the fundamental light of FWHM$\sim$ 1.2 GHz.

The absolute laser wavelength at each frequency step was recorded with a Lambdameter LM-007, calibrated by a frequency-stabilized He-Ne laser. An average energy per pulse of $\sim$300~$\mu J$ was found sufficient to saturate the first step transition, while for the ionization step to the continuum an energy per pulse of up to 20~mJ could not provide full saturation for this transition.

In a reference cell near the laser table an atomic beam of naturally occurring  $^{107,109}$Ag isotopes was produced in a graphite crucible. A small fraction of the laser beams was directed to the reference cell, where ionization of the atomic beam was carried out in a 90$^\circ$ cross-beam geometry. On the other hand, by resistively heating a silver filament in the gas cell an atomic vapor of stable isotopes was readily available. Ionization of these beams was used for on-line tuning purposes and for monitoring systematic effects as, e.g., fluctuations of the laser power and frequency shifts.

\section{Data Analysis}
\begin{figure}[h]
\begin{center}
\includegraphics[scale=0.65]{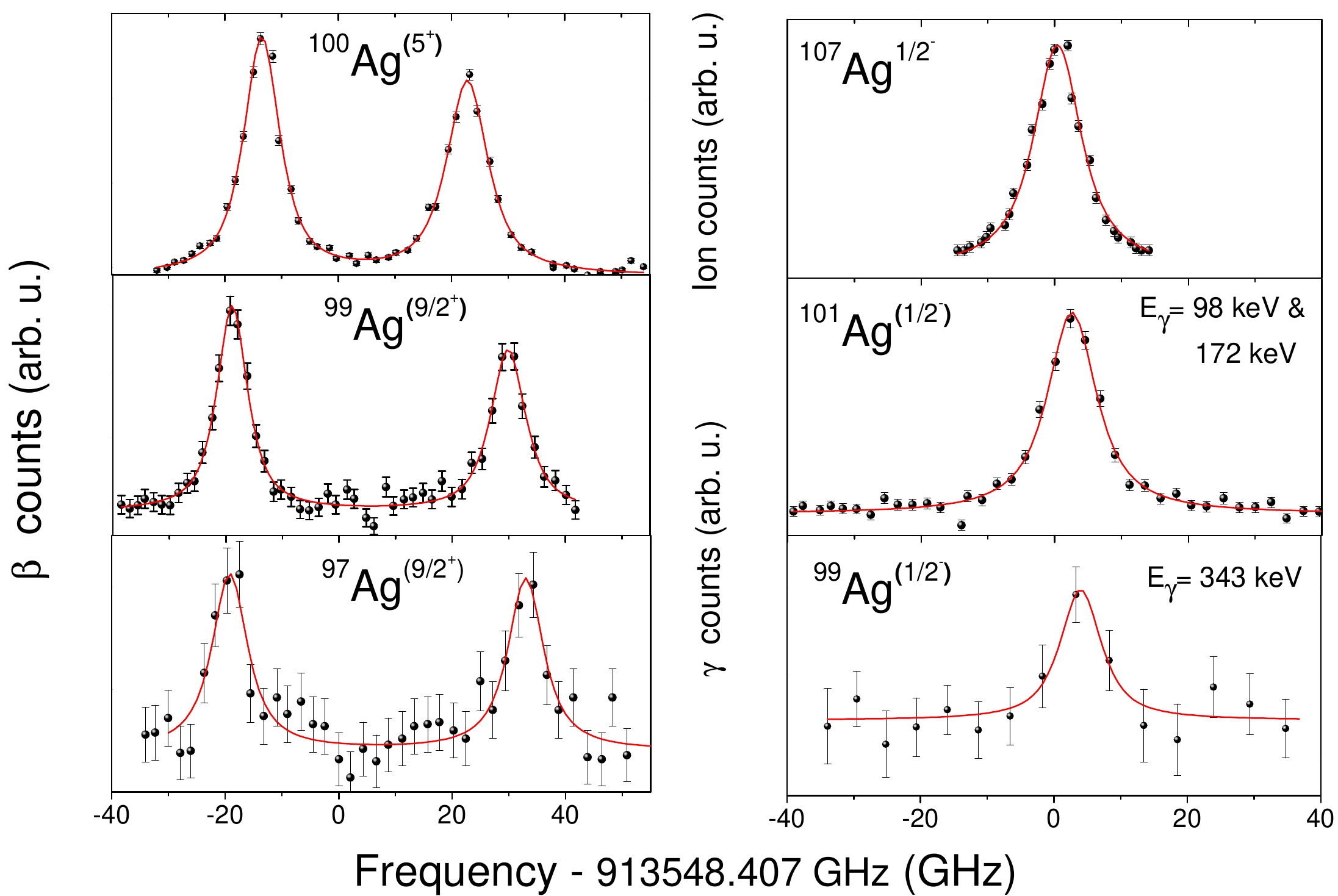} \caption{Left panel: detected $\beta$-counts as a function of the scanning laser frequency. Right panel: resonance of the stable $^{107}$Ag isotope (at a buffer gas pressure of p$=$ 520 mbar) used to study different systematic effects and the resonances corresponding to the isomeric states obtained by gating on the specified $\gamma$-transitions. Fits of a single- or double-peak Voigt profile to the data points are shown.}\label{res:odd}
\end{center}
\end{figure}

Laser frequency scans were performed and the $\beta$- and $\gamma$-particle count rates monitored simultaneously in each frequency shot. As a result, a number of $\beta$ and $\gamma$ resonances were obtained for $^{98-101}$Ag, while in the case of $^{97}$Ag the yield was such that only $\beta$ data was possible. As an example, spectra for the silver isotopes using $\beta$ and $\gamma$ decays are shown in Fig. \ref{res:odd}. The sum of two laser scans are shown in all panels, while data analysis was performed on each scan individually. An offset corresponding to the center of gravity (CoG) of the scanned transition in the stable $^{109}$Ag isotope \cite{Bad04} is subtracted from the laser frequency to indicate the frequency detuning.

Fitting of the obtained resonance lineshapes was carried out using Voigt profiles, which comprise the convolution of both Gaussian and Lorentzian components in the fits. Based on measurements of the stable $^{107}$Ag isotope in the reference cell, a Gaussian FWHM of 1.90(7)~GHz was assigned to the laser linewidth, neglecting a $\sim$10\% residual Doppler broadening effect due to the atomic beam divergence. In the gas cell, the Doppler broadening is governed by the gas temperature, which we assume to be 350(25) K based on off-line measurements and on an estimate of the heating power transferred by the incident primary beam. The resulting Doppler width can thus be estimated to be 1.18(5)~GHz for this transition. These two contributions, laser linewidth and Doppler broadening, resulted in a total Gaussian FWHM of 2.5(3) GHz that was used as a fixed parameter in all fits. On the other hand, the width of the Lorentzian distribution, including the laser power- and pressure-broadening contributions (neglecting the broadening due to the natural linewidth of $\sim$20~MHz for the 5s$^2$ S$_{1/2}$ $\rightarrow$ 5p$^2$ P$_{3/2}$ transition), was allowed to remain free.

Measuring the resonance width of $^{107}$Ag for different pressures (at the same laser power) in the gas cell, a pressure-broadening coefficient $\gamma_{coll}=$ 12(2) MHz/mbar could be obtained. Using this value, a contribution of the power broadening to the Lorentzian widths of up to 4 GHz, mostly present in the measurements of $^{98,97}$Ag, was determined. This resulted in a total FWHM of about 10 GHz.

%which resulted in a total Lorentzian FWHM of about 8 GHz.

Although each of the resonance peaks in our data consists of three hyperfine transitions (see top of Fig. \ref{ion:sch}), the small splitting of the atomic excited states and the magnitude of the pressure broadening led to non-resolved structures. Fits with both triple- and single-peak of the unresolved structures were performed. The obtained peak centroids coincided in both cases and demonstrated our insensitivity to the substructure of the observed broad resonances.
As a result, the resonance line-shape of the low-spin isomers and of the stable isotope $^{107}$Ag were fitted using a $\chi^{2}$ minimization of a single-peak Voigt profile (right panel in Fig. \ref{res:odd}), while a double-peak Voigt profile was used for the high-spin ground states (left panel in Fig. \ref{res:odd}).

From each fit to the data exhibiting a double-peak resonance, the separation $\Delta$$E$ between the two centroids was determined. As the observable $\Delta$$E$ depends on both the ground- and excited-state hyperfine parameters, to be able to extract the  ground-state splitting $\Delta\nu$ and deduce the hyperfine parameter A(4d$^{10}$5s$^{2}$ $^{2}$S$_{1/2}$) of the ground state, the hyperfine parameter of the excited state A(4d$^{10}$5s$^{2}$ $^{2}$P$_{3/2})$ had to be considered as well. As the latter could not be resolved in our measurements, the well-known ratio of magnetic hyperfine parameters A(4d$^{10}$5s$^{2}$ $^{2}$P$_{3/2})$/A(4d$^{10}$5s$^{2}$ $^{2}$S$_{1/2}$)$=$0.0186(4) from the stable isotope $^{109}$Ag \cite{Dah67,Car90} was used to deduce the ground-state splitting of the isotopes investigated in this work. A variation of this ratio along the isotopic chain caused by the presence of hyperfine anomaly (see below) was evaluated and found to be negligible to within our experimental uncertainty.

As the intensities of the resonance peaks did not stay constant throughout the measurements, calculations of the Racah intensities neglecting optical pumping effects were used instead to deduce the center of gravity of the atomic transition for each isotope. In this way, the isotope shift (IS) could be determined. Variations of the relative intensities of about 50\%  between  different scans for a given isotope were observed, while the averaged value of the relative intensities deviated from those calculated in the range of 5\% (for $^{100}$Ag) to 15\% (for $^{99}$Ag).

An important source of statistical uncertainty was attributed to the multi-mode nature of the light used in the scanning (first step) laser. The mode behavior changed arbitrarily during laser scans, leading to a random scattering of the wavelength readout. Measurements over a long period of 24 hours of a natural admixture of stable silver isotopes $^{107,109}$Ag in the reference cell showed frequency fluctuations with respect to the literature value \cite{Pic01}. Assuming a most likely unfavorable situation, an error to the wavelength meter readout of 150 MHz was assigned to the frequency values and added in quadrature to the statistical uncertainty from the data analysis.

Besides the presence of statistical fluctuations in the experimental data, systematic effects were evaluated and accounted for in the final frequency values.
Frequencies were corrected after calibration of the wavelength meter. The mean value obtained from the readout on the He-Ne laser before and after each laser scan was used as calibration value. Moreover, a systematic frequency shift caused by atomic collisions with the buffer gas atoms was evaluated and corrected in the scanned laser frequencies. A linear fit of the resonance centroids obtained for $^{107}$Ag in the gas cell at different buffer gas pressures raging from 125 to 520 mbar were used to deduce a pressure-shift coefficient of $\gamma_{sh}=$ -3.7(4) MHz/mbar. The extrapolated value of the resonance centroids to a pressure of p$=$0 mbar after calibration of the wavelength meter resulted in a CoG$_{GC}^{107}=$913,548,94(16)(11) MHz, which is in good agreement with the literature value CoG$_{lit.}^{107}=$913,548,883(30) MHz \cite{Uhl00}. An error arising from the pressure shift determination was assigned to the deduced CoG in each isotope.

A final error, including statistical and systematic uncertainties added in quadrature, was assigned to the weighted mean of the resonance centroids for a given isotope.  Internal versus external uncertainty consistencies were compared and their ratio was found to be close to unity, indicating that only statistical fluctuations were present in the experimental data \cite{Bir32}. The larger value of the two uncertainties was taken as the final error.
%The pair of centroids obtained in each frequency scan of a given isotope were then used to obtain a weighted mean of the observable $\Delta E$.

In Tab. \ref{tab:1} the HFS parameters obtained in the data analysis are listed. Statistical and systematic (when present) uncertainties are given separately. Note that from the considered sources of systematic uncertainty only that associated with the determination of the pressure shift, which cancels for the observable $\Delta$$E$, is shown in brackets. The calibration of the wavelength meter by the He-Ne laser and the pressure shift were directly corrected in the frequency values.
%\begin{sidewaystable}
\begin{table*}[h]
\caption{Summary of the results obtained from spectroscopic studies of the $5s ^2S_{1/2}\rightarrow5p ^2P_{3/2}$ transition ordered by mass number and nuclear state. The nuclear spin I and the ground- and excited-state hyperfine parameters are given from the second to the fourth columns, respectively. In the last three columns the magnetic moments, isotope or isomer shifts, and the changes in the mean-charge radii are listed. Statistical and systematic (if any) contributions to the errors are given separately (see text for details).} \label{tab:1}
\centering
\begin{threeparttable}
\begin{tabular}{lccccr@{.}lr@{.}l}
\toprule
Isotope & I & A$(^2S_{1/2})$ & A$(^2P_{3/2})\tnote{a}$ & $\mu_{exp}$ &\multicolumn{2}{c}{$\delta\nu^{A,109}$} & \multicolumn{2}{c}{$\delta\langle r^2\rangle^{A,109}$} \\ \midrule
  & &(GHz)& (GHz) &($\mu_N$) & \multicolumn{2}{c}{(GHz)} & \multicolumn{2}{c}{(fm$^2$)}\\ \hline		
97&(9/2)&10.6(2)&\,\,\,0.197(11)& 6.13(12)&\,\,\,\,\,-4&67(66)(8)&\,\,\,\,\,\,\,\,1&29(20)(9)\\
\multirow{2}{*}{98}&(6)&6.02(9)&0.112(6)&4.64(7)&-3&58(47)(8)&1&01(16)(7)\\
                    &(5)&\,\,\,7.12(11)&0.132(7)&4.57(7)&-3&34(47)(8)&0&95(16)(7)\\
99&(9/2)&\,\,10.05(5)&\,\,\,0.187(10)&5.81(3)&-3&21(25)(11)&0&91(12)(7)\\
99$^{(i.s.)}$&(1/2)&&&&-4&00(94)(11)&1&10(25)(8)\\
100&(5)&6.81(4)&\,0.126(7)&4.37(3)&-2&95(22)(11)&0&83(10)(6)\\
101&9/2&9.64(6)&\,\,\,\,0.179(10)&5.57(4)&-2&48(26)(11)&0&70(10)(6)\\
101$^{(i.s.)}$&(1/2)&&&&-2&40(21)(11)&0&68(9)(5)\\
107&1/2&&&&-0&53(16)(11)&0&15(4)(3)\\ \bottomrule       	
\end{tabular}
\begin{tablenotes}
\footnotesize
\item[a] Obtained from the HFS of $^{109}$Ag.
\end{tablenotes}
\end{threeparttable}
\end{table*}
%\end{sidewaystable}

Spectroscopic information on the known isomeric states in $^{99,101}$Ag could also be obtained. Coupling of the used ionization scheme to the nuclear spin of 1$/$2 results in four sub-states with quantum numbers F$=$(0,1) \& (1,2), which produces a hyperfine structure with three components. By monitoring the amplitude of the $\gamma$ lines from the isomeric decay as a function of the laser frequency, the isomeric transitions were observed (see Fig. \ref{res:odd}).

As in the case of the HFS for the larger nuclear spins, the individual (three) hyperfine components  could not be observed owing to the lack of spectral resolution that for these isomeric states did not allow to resolve the small ground- and excited-state hyperfine parameters. For instance, in the $^{107}$Ag isotope (I$=$1/2) one finds a value of A($^2$S$_{1/2}$)$\sim$ 1.7 GHz \cite{Dah67} and A($^2$P$_{3/2}$)$\sim$ 30 MHz \cite{Car90}. Similar values of the HFS parameters for $^{99,101}$Ag can be expected according to the observed spectral linewidths, comparable to that in $^{107}$Ag.
Nevertheless, the centroid of these resonances was used in a good approximation (as confirmed in the measurements of the pressure shift using $^{107}$Ag) as the CoG of the transition in order to determine the isomer shifts.

\section{Results}
Although the splitting of the excited state and the collapsed hyperfine structure of the isomeric states could not be resolved, useful information was extracted from both the measured ground-state splitting and the centers of gravity, which allow one deducing the magnetic dipole moments and the changes in the mean-charge radii, respectively.

\subsection{Magnetic Dipole Moments and g factors}
The magnetic dipole moment $\mu$ is related to the hyperfine splitting parameter A by $\mbox{A}=\mu \cdot B_0/(I \cdot J)$ \cite{Neu06},
%\begin{equation}\label{mu:a}
%    A=\frac{\mu \cdot H_0}{I \cdot J}\,\, ,
%\end{equation}
where B$_0$ is the magnetic field generated by the motion of the electrons, orbiting with a total angular momentum J, at the position of a nucleus with spin I. B$_0$ is specific to the atomic state considered and therefore isotopically independent. Thus, assuming the nuclear spin of the level concerned and using the hyperfine parameters determined experimentally along with the accurate data from a reference measurement, the magnetic moment of the isotopes under investigation can be usually obtained by a scaling relation such as
\begin{equation}\label{mu:b}
    \mu^{exp}=(1+ ^{ref}\Delta^{exp})\cdot\frac{A^{exp}(4d^{10}\, 5s^{2}\, ^{2}S_{1/2})\cdot I^{exp}}{A^{ref}(4d^{10}\, 5s^{2}\, ^{2}S_{1/2})\cdot I^{ref}}\cdot\mu^{ref}\,\,.
\end{equation}

\subsubsection{HFS anomaly}
The term $^{ref}\Delta^{exp}$ in Eqn. \ref{mu:b} refers to the differential hyperfine anomaly and accounts for the different distribution of the nuclear magnetization in the volume of the reference nucleus and in that of the isotope of interest. In the silver isotopes one finds that while the hyperfine anomaly between similar nuclei--spin, nucleon configuration and magnetic moments--are often negligible within the experimental accuracy \cite{Bla66, Fis75}, relatively large values are measured between unlike nuclear structures caused by a  quite different origin of nuclear magnetism in the two nuclei involved, as e.g  $^{109}\Delta^{110m}$($^2$S$_{1/2}$)$=$2.47(12)$\cdot$10$^{-2}$ \cite{Sch67} and $^{107}\Delta^{103}$($^2$S$_{1/2}$)$=$-3.1(1.7)$\cdot$10$^{-2}$ \cite{Wan70}. Furthermore, changes in the nuclear potential experienced by the electrons due to the extended (charge and current distribution) nucleus are more significant in configurations containing unpaired electrons in the S$_{1/2}$ ground-state, and to a lesser extent in the P state, because of their non-zero charge densities at the nucleus \cite{Bue84}.

As the spin and magnetic moment of the reference nucleus ($^{109}$Ag) are very different to those of the isotopes investigated, and the atomic transition whose HFS wants to be determined involves an unpaired electron in the $^2$S$_{1/2}$ ground state, the hyperfine anomaly was included in our analysis of the magnetic moments.

The differential hyperfine anomaly between two isotopes can be written as
$^{1}\Delta^{2}=(A_1\cdot I_1\cdot\mu_2/(A_2\cdot I_2\cdot\mu_1))-1\approx \epsilon_1-\epsilon_2$,
%\begin{equation*}\label{anom:aly}
%    ^{1}\Delta^{2}=\frac{A_2}{A_1}\frac{I_2}{I_1}\frac{\mu_1}{\mu_2}-1 \approx \epsilon_1-\epsilon_2\,\, ,
%\end{equation*}
where $\epsilon$ refers to the hyperfine anomaly or Bohr-Weisskopf effect \cite{BW49} of each isotope. To estimate the values of $\epsilon$ for the isotopes investigated in our experiments, the semiempirical  approach of Moskowitz and Lombardi  $\epsilon=\alpha/\mu$ was used \cite{Mos73}. With such a simple relation these authors obtained good agreement with the available experimental values in the mercury isotopes. This equation has also been recently  found useful to estimate the hyperfine anomaly in the neighboring cadmium isotopes \cite{Yor13}.

Using the available experimental data from the literature for the silver isotopes with atomic mass A$=$ 103, 107 and 109 \cite{Wan70, Car90, Dah67} one obtains a value of $\alpha=$3.61(2)$\cdot$10$^{-3}\mu_N$. This value results in differential hyperfine anomalies $\Delta$($^2$S$_{1/2}$) between these isotopes that are in good agreement with those determined experimentally \cite{Bla66, Wan70}. For the odd-even isotopes with atomic masses A$=$ 97, 99, 101 and spin I$=$9$/$2 measured in our experiments one obtains a value of $^{109}\Delta^{A}$($^2$S$_{1/2}$)$\sim$~-2.8~\%. For the odd-odd isotopes $^{98}$Ag and $^{100}$Ag the calculations give also a similar result. This latter case is to be compared with the precise experimental value of $^{109}\Delta^{110m}$($^2$S$_{1/2}$)= 2.47(12) \% \cite{Sch67} for which there exists a similar spin relation (I($^{110m}$Ag)$=$ 6) between the nuclei involved.

The experimental values of the magnetic moments and their corresponding error bars (see Table \ref{tab:1}) were calculated from Eqn. \ref{mu:b} using the measured hyperfine parameters A($^2$S$_{1/2}$) and including the correction for the presence of hyperfine anomaly.

\subsection*{The odd-odd Ag isotopes}
\begin{figure}[h]
\begin{center}
\includegraphics[scale=0.5]{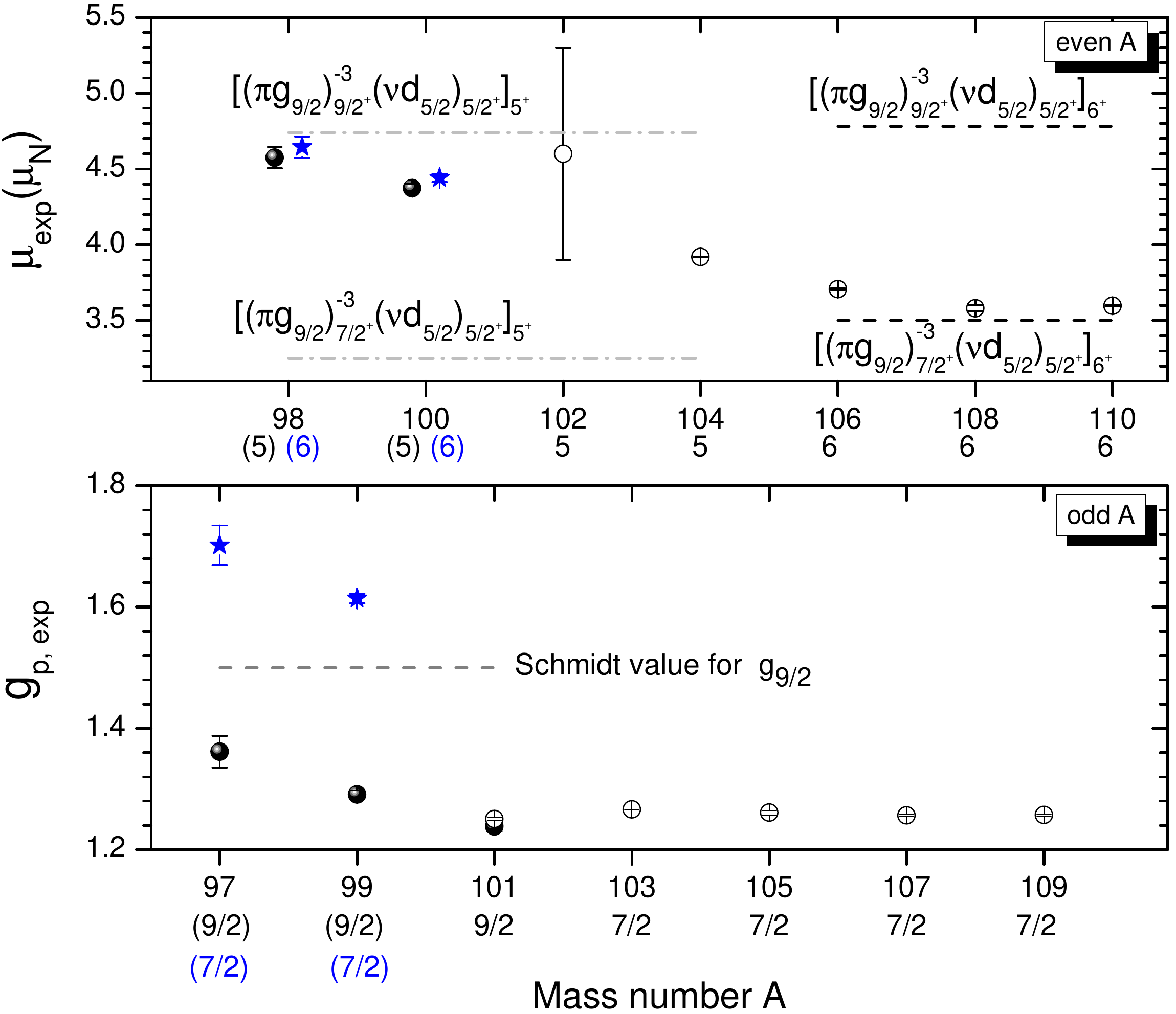} \caption{Top panel: experimental magnetic moments from this work assuming a 5$^+$ (filled circles) and 6$^{+}$ (stars) spin together with previous results from literature values (open circles) for the even-A isotopes. Theoretical magnetic moments producing a spin 5$^{+}$ (dot-dashed grey line) and 6$^{+}$ (dashed black line) are indicated. Bottom panel: filled circles and stars show the derived experimental proton g-factors for the odd-A isotopes measured in this work assuming a spin of 9$/$2$^+$ and 7$/$2$^+$, respectively. Previous experimental results are indicated with open circles. See text for details.}\label{mu:gp}
\end{center}
\end{figure}

In the upper panel of Fig. \ref{mu:gp} the evolution of the experimental magnetic moments of the high-spin  $\beta$-decaying states in the odd-odd Ag--for literature values see V. V. Golovko et al. \cite{Gol10}--is given from A$=$110 down to A$=$98 together with the spin assignments found in the literature. The calculated magnetic moments  using the additivity rule \cite{Bre60} with effective gyromagnetic ratios g$_l^{mod}$($\pi$)$=$+1.1, g$_l^{mod}$($\nu$)$=$-0.05 and g$_s^{mod}=$ 0.7g$_s^{free}$ (see also \cite{Gol10}) are shown in the figure for three protons in the $g_{9/2}$ orbit coupled to $d_{5/2}$ neutrons to produce a spin 5$^{+}$ (dot-dashed gray line) and 6$^{+}$ (dashed black line) for J$=$\,j-1$=$\,7/2 (lower lines) or J$=$j$=$\,9/2 (upper lines). For the $^{106,108,110}$Ag isotopes the magnetic moment of the  6$^+$  $\beta$-decaying isomer can be explained by the parallel coupling of the $(\pi\mbox{g}_{9/2})^{-3}_{{7/2}^+}$ proton group and the $\nu$d$_{5/2}$ neutron hole. The $^{102,104}$Ag have been described as the coupling of the $\nu$d$_{5/2}$ neutron configuration with a mixed ($\pi$g$_{9/2}$)$^{-3}_{{7/2}^+}$ -($\pi$g$_{9/2}$)$^{-3}_{{9/2}^+}$ configuration \cite{Gol10,Ame61}. Our measurements on $^{98,100}$Ag show a further evolution  towards a pure ($\pi$g$_{9/2}$)$^{-3}_{{9/2}^+}$ $\otimes$ $\nu$d$_{5/2}$ configuration. Owing to the limited spectral resolution of the method employed it is not possible to make a firm spin assignment (see the open star in Fig. \ref{mu:gp}, top panel).

\subsection*{The odd Ag isotopes}
The evolution of the experimental g-factors from A$=$109 down to A$=$97 (A$=$109-107 from R. Eder et al. \cite{Ede85} and A$=$105-101 from U. Dinger et al. \cite{Din89}) is shown in the lower panel of Fig. \ref{mu:gp} for the high-spin  $\beta$-decaying states. The experimental g-factors found in the literature are remarkably constant although the ground-state configuration changes in $^{101}$Ag from ($\pi$g$_{9/2}$)$^{-3}_{{7/2}^+}$ to ($\pi$g$_{9/2}$)$^{-3}_{{9/2}^+}$. This is in good agreement with the predictions of V. Paar \cite{Paa73}. Our result for the nuclear magnetic moment of  $^{101}$Ag (5.57(4)~$\mu_{N}$) is in good agreement with the tabulated value 5.7(4)~$\mu_N$ found in the literature \cite{Sto05} and also with the more precise value of 5.627(11)~$\mu_N$ measured by U. Dinger et al.

Our new nuclear magnetic moments for $^{97,99}$Ag are only giving g-factors consistent with the observed constant trend when a spin of 9$/$2 is assumed (see the star symbols in the bottom panel of Fig. \ref{mu:gp} corresponding to the g-factor using 7$/$2 as spin value). The spin 9/2 in these nuclei is expected to originate from seniority one three-hole configurations ($\pi$g$_{9/2}$)$^{-3}_{{9/2}^+}$ as in the case of $^{101}$Ag \cite{Van83}. Furthermore, the Schmidt value for the gyromagnetic ratio of a single particle g$_{9/2}$ is 1.5 and the g-factors assuming a spin 7$/$2 would lay above it. The fact that for $^{97}$Ag the g-factor approaches the Schmidt value is an indication of the enhanced single particle behavior at the N$=$50 shell closure. It is worth noticing also that the g-factor of the N$=$50 isotone $^{93}$Tc (I$=$ 1.405(14)) \cite{Hin95} is within the same range. Under these premises, one can tentatively assign a value of 9/2 as the most likely spin for $^{99,97}$Ag.

The collapsed structure in the laser resonance for the low-spin isomer in $^{99,101}$Ag indicates that the magnetic moment of these states is very small in accordance with a spin 1$/$2. Combined with the decay information of $^{99m}$Ag from M. Huyse et al. \cite{Huy81} and $^{101m}$Ag from D.J. Hnatowich et al. \cite{Hna70} and the suggested ground-state spin (see above), the intermediate level can be spin assigned as 7$/$2$^+$ .

\subsection{Isotope Shifts and Mean-Square Charge Radii}

The isotope shift (IS) between two isotopes with mass numbers A and A$'$ is manifested as a slight frequency variation of the center of gravity $\nu$ of the atomic spectral components for a given transition. The IS is usually expressed as \cite{Klaus13}
\begin{equation}\label{IS}
    \delta \nu^{A,A'}= \nu^{A'}-\nu^{A}=\delta \nu_{NMS}^{A,A'}+\delta \nu_{SMS}^{A,A'}+\delta \nu_{FS}^{A,A'}
\end{equation}
with the mass shift (MS), separated in normal (NMS) and specific (SMS) mass shift components, that are due to the finite mass and non-zero size of the nucleus, and the field shift (FS) accounting for the different interaction undergone by the shell electrons with the nuclear isotopic contents. While the calculation of the NMS is trivial,
%\begin{equation}\label{NMS}
%    \delta \nu_{NMS}^{A,A'}=\frac{\nu}{1836.1}\cdot\frac{A'-A}{A'\cdot A}\,\, ,
%\end{equation}
the SMS is in general difficult to determine \cite{Che12}. Here we will use the SMS estimate for ns--np transitions suggested by P. Aufmuth et al. \cite{Auf87} of $\delta\nu_{SMS}^{A,A'}= 0.3(9)\cdot\delta \nu_{NMS}^{A,A'}$.
%\begin{equation}\label{SMS}
%    \delta\nu_{SMS}^{A,A'}= 0.3(9)\cdot\delta \nu_{NMS}^{A,A'}\,\, .
%\end{equation}
The FS contains the isotopic variation of the mean-square charge radius and can be written as
$\delta \nu_{FS}^{A,A'}=F\cdot\lambda^{AA'}=F\cdot(\delta\langle r^2\rangle^{AA'}+T)$
%\begin{equation}\label{FS}
%    \delta \nu_{FS}^{A,A'}=F\cdot\lambda^{AA'}=F\cdot(\delta\langle r^2\rangle^{AA'}+T)\,\, ,
%\end{equation}
with the convention that $\delta\langle r^2\rangle^{AA'}=\langle r^2\rangle^{A'}-\langle r^2\rangle^A$. The term T accounts for higher radial moments and was parameterized by E. Seltzer \cite{Sel69} as  $T=(C_2/C_1)\delta\langle r^4\rangle^{AA'}+(C_3/C_1)\delta\langle r^6\rangle^{AA'}+...$, with the coefficient $C_i$ later on recalculated by Blundell et al. \cite{Blu87}. By modeling the nucleus as a homogeneously charged sphere of radius R$=1.2A^{1/3}$ and calculating $\delta\langle r^i\rangle$ one obtains for T a value of 0.976 \cite{Din89}.
F is the electronic factor, which is proportional to the change of the electronic charge density at the nucleus in the transition under investigation. G. Fricke et al. \cite{Fri95} give a detailed explanation of the different procedures to determine it. Here we will used the semi-empirical estimation of F$=\,$-4300(300)~MHz$/$fm$^2$ from Blundell et al. for the 5s$\rightarrow$5p transition at 328 nm obtained from the 5s hyperfine structure \cite{Dah67}.
%For a summary of the electronic factors for the 328, 547.7, and 1937 nm lines in atomic silver obtained by different \emph{ab initio} and  semi-empirical procedures we refer the reader to the work of U. Dinger et al. \cite{Din89}.

The results of the measured isotope shifts and those from the analysis of the mean-square charge radii, taking into account the aforementioned estimates, are listed in the last two columns of Tab. \ref{tab:1}. The error of the field shift constant F is included in the systematic uncertainty of $\delta\langle r^2\rangle$.
\begin{figure}
\begin{center}
\includegraphics[scale=0.35]{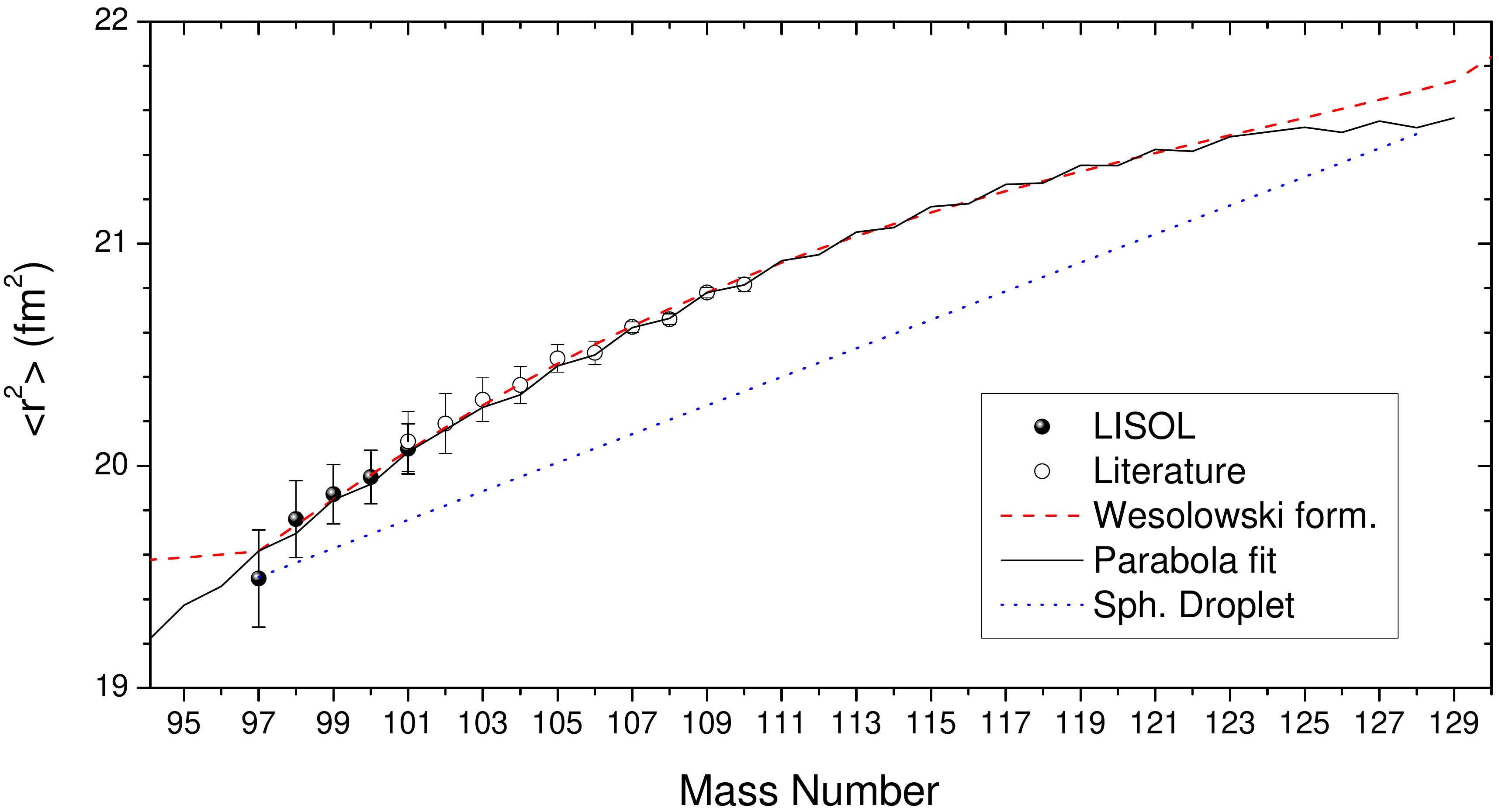} \caption{Mean-square charge radii of the silver isotopes from A=110 to A=97 (shell closure). The LISOL values are shown with filled circles along with the total (statistical plus systematic) errors, while with open circles are the values corresponding to previous measurements from \cite{Din89}. The dashed line represents the results of the semi-empirical formula of Weso{\l}owski \cite{Wes85}, whereas the solid line is the best fit of the parabolic polynomial described in \cite{Ott89}. The prediction of the droplet model, as given in \cite{Din89}, is shown with a dotted line.}\label{fig:8}
\end{center}
\end{figure}

Figure \ref{fig:8} illustrates the mean-square charge radii for the isotopes studied in this work together with the data for the heavier silver isotopes obtained by U. Dinger et al. \cite{Din89}. In the same fashion as done by these authors, the results for the isotopic chain, now complete down to the N=50 shell closure, are presented along with the predictions of the semi-empirical formula (dashed line) of E. Weso{\l}owski \cite{Wes85}. The absolute values of the radii were extracted by using the estimation of $\langle$r$^{2}$($^{109}$Ag)$\rangle$ predicted by Weso{\l}owski's formula. A parabola containing an odd-even term was fitted to the data points and is also shown in the figure. Parabolic shapes of $\delta\langle r^2\rangle$  are often found in regions where shape transitions evolve smoothly \cite{Ott89}. Such a parabolic behavior has been observed in the neighboring chains of Cd \cite{Buc87}, In \cite{Ebe87}, and Sn \cite{Eber87} isotopes. The curvature of the parabola is related with the quadrupole contribution in a core polarization model as suggested by I. Talmi \cite{Tal84}. Unfortunately, the large error bars of our data do not allow extracting any conclusion on the dynamics of the nuclear deformation from the IS data. The large uncertainties also mask the presence in our data of the odd-even staggering, clearly observed in the heavier silver isotopes. Our results show a trend towards spherical nuclear shape for the nuclei in the proximity and at the N=50 shell closure consistent with the predictions of the droplet model \cite{Mye83} for non-deformed ($\beta$=0) nuclei (dotted line). The predicted linear behavior, as given in \cite{Din89}, results from the values of the radii obtained when the droplet model is used for the spherical part of the two-parameter (spherical and deformation term) model formula \cite{Eber87, Ott89}.

\section{Summary and Outlook}

Laser ionization spectroscopy of radioisotopes produced in heavy-ion fusion evaporation reactions and stopped in a buffer-gas cell have been successfully carried out. Magnetic dipole moments and mean-square charge radii from $^{101}$Ag down to $^{97}$Ag at the N$=$50 shell closure were obtained. Good agreement was found with the results from the literature for the isotope $^{101}$Ag, the only one--from those presented in this article--measured previously.

The new results confirm the importance of the N$=$50 shell closure  and can be compared to more elaborated shell-model calculations around $^{100}$Sn. The limited spectral resolution, however, prevented us from making firm spin assignments and reducing the uncertainties, especially on the mean-charge radii. Moreover, to reach further towards the interesting region of the N$=$Z isotopes (e.g. $^{94}$Ag), more intense heavy-ion beams are essential.

A combination of the recently developed in-gas jet laser ionization spectroscopy technique \cite{Yur13} with the new heavy-ion accelerator facility of SPIRAL2 (GANIL) will allow creating the necessary conditions in order for the studies in the N=Z region to be performed \cite{Fer13}.

\section*{Acknowledgments}
We thank the CRC team, Louvain-La-Neuve (Belgium), for providing the primary beams, and G. Neyens, J. Papuga, and S. Raeder for fruitful discussions.
This work was supported by: FWO-Vlaanderen (Belgium), GOA/2010/010 (BOF KULeuven), the IAP Belgian Science Policy (BriX network P7/12), the European Commission within the Seventh Framework Programme through I3-ENSAR (contract No. RII3-CT-2010-262010), and a grant from the European Research Council (ERC-2011-AdG-291561-HELIOS).

\end{document}